\documentclass[letterpaper, 10 pt,conference]{cssconf}  

\IEEEoverridecommandlockouts
\usepackage{amssymb}
\usepackage{amsmath}
\usepackage{graphicx}
\usepackage{cite}
\usepackage{multicol}
\usepackage{enumerate}
\usepackage{lipsum} 
\usepackage[capitalise]{cleveref}
\usepackage{tikz}
\usetikzlibrary{patterns}
\usepackage{accents}
\usepackage{tikz}

\newtheorem{proposition}{Proposition}
\newtheorem{remark}{Remark}
\newtheorem{theorem}{Theorem}
\newtheorem{definition}{Definition}
\newtheorem{corollary}{Corollary}
\newtheorem{lemma}{Lemma}

\DeclareMathOperator{\Equaldef}{\overset{def}{=}}

\title{\LARGE \bf Global games with Poisson observations: Bio-inspired distributed coordination of multi-agent systems}
\author{Marcos M. Vasconcelos
\thanks{The work of M. M. Vasconcelos is supported by funds from the Commonwealth Cyber Initiative (CCI). The author is with the Commonwealth Cyber Initiative and the Bradley Department of Electrical Engineering, Virginia Tech. 900 North Glebe Rd, Arlington, Virginia, 22203 - USA. E-mail: \texttt{marcosv@vt.edu}.}
}
\begin{document}

\maketitle

\begin{abstract}
Global games are a class of incomplete information games where the payoffs exhibit strategic complementarity leading to an incentive for the agents to coordinate their actions. Such games have been used to model scenarios in many socioeconomic phenomena, where the private signals available to the agents are typically assumed to be Gaussian. We study an instance of a global game where the agents observe Poisson random variables, which are inspired by applications in microbiology where information signals are disseminated via discrete molecular signals rather than continuous. Although this observation model violates the essential technical assumptions present in the Gaussian case, we present preliminary results on the existence of Bayesian Nash equilibria in pure threshold policies in two variants of the underlying random state-of-the-world: an arbitrarily distributed discrete binary state and a continuous state with uniform distribution.
\end{abstract}

\maketitle

\section{Introduction}

Global games constitute a class of games with incomplete information where the agents exhibit strategic complementarity and have incentive to coordinate collective behavior \cite{Carlsson1993GlobalSelection}. The framework introduced in \cite{Carlsson1993GlobalSelection} has been used to model strategic behavior in socioeconomic applications ranging from bank runs to political revolutions \cite[and references therein]{Morris2003GlobalApplications}. Interestingly, global games have also been used to model engineered systems such as decentralized mode selection in sensor networks \cite{Krishnamurthy2008DecentralizedGames,krishnamurthy2011networks} and task-allocation in robotic networks \cite{Kanakia2016ModelingGame}. Therefore, the framework of global games has been extremely versatile to model applications in which the agents are incentivized to coordinate their actions to accomplish collective tasks.

Multi-agent coordination is also prevalent in microorganisms: It has been widely accepted that essentially every species of bacteria coordinate density-dependent collective behavior via a mechanism 
known as \textit{quorum sensing} (QS) \cite{Miller:2001,Darch2012Density-dependentPopulations,Boedicker2015MicrobialSensing}. Although, the biochemical and genetic pathways involved in QS are well-documented in the literature, its rigorous mathematical underpinnings have received much less attention. One notable exception is the work in \cite{Michelusi2016QueuingCommunities}, which models QS using queuing theory, but is limited by the lack of decision-making abilities of the agents, and turns out to be equivalent to a model based on deterministic differential equations. The work in \cite{Vasconcelos2018BacterialSystem} introduced a mathematical model for QS in which the agents make imperfect observations about the state-of-the-world and make decisions with the goal of maximizing a payoff function consisting of a public benefit and a local cost terms. However, the work in \cite{Vasconcelos2018BacterialSystem} assumes that all agents use identical threshold policies and, therefore, are already acting in a coordinated fashion whose optimization is carried out by the so-called \textit{social planner}.

Inspired by the problem of coordinating collective behavior in bacterial QS, we study the problem of multi-agent coordination using a global game formulation where the agents sense the state-of-the-world via Poisson observations. The motivation for assuming Poisson observations is that in microbiology, the cells interact with each other and sense their environment employing molecular signals, which are inherently discrete. Therefore, the Gaussian observation model so prevalent in the global game literature is not appropriate. Furthermore, with the lack of a continuous and well-behaved Gaussian model, the existing arguments for the existence of Bayesian-Nash equilibria (BNE) \cite{Fudenberg1991Game1991} break down and are no longer available. On the other hand, the bacterial QS system is characterized by threshold policies that control homeostatic cellular behavior despite the presence of discrete of molecular communications. Therefore, if bacterial QS is to be modeled using a global game model, a BNE within the class of threshold strategies should exist. Unlike Gaussian signals/random variables, Poisson random variables do not admit an asymptotic analysis in the noiseless/high SNR regime.

Our goal is to establish conditions under which threshold policies constitute an invariant class of policies in the following sense: the best-response strategy to an arbitrary threshold strategy profile for the opponents in the game is itself a threshold policy. We provide a proof of this result in two cases: when the state-of-the-world is  binary with an arbitrary distribution, and in the case of a continuous uniform distribution. More importantly, we provide sufficient conditions for the existence of a BNE in pure threshold strategies, in the absence of continuity of the best-response mapping, and of a finite policy space for the agents.

The proofs rely on a simplified structure of the payoff function of the agents, which capture \textit{strategic complementarity}, i.e., that taking a particular action becomes more attractive the more other agents engage in the same action \cite{chamley2004rational,Dahleh:2016}. We follow simple arguments and techniques present in \cite{Touri2014GlobalInformation,mahdavifar2017global,Kanakia2016ModelingGame,Dahleh:2016}. However, the Poisson distribution plays a major role in all of our arguments, often leading to expressions resembling combinatorics, which seem to be documented in the global game literature for the first time here.

\section{System Model}

Consider a system with $N$ agents, and denote the collection of all agents by $[N]\Equaldef \{1,2,\cdots,N\}$. We assume that total number of agents in the system is common knowledge among all agents. Each agent can take a binary action $a_i\in\{0,1\}$, which represents the decision to engage in a free or costly behavior, respectively. Let $a_i=0$ denote the $i$-th agent's decision to \textit{not activate}, and $a_i=1$ denote its decision to \textit{activate}. For example, in bacterial quorum sensing, this action might correspond to the decision to begin the production of an energetically costly enzyme or to express a gene responsible for so-called virulence attacks.

The payoff of not activating is normalized to zero, whereas the payoff of activating depends on the number of other agents who also decide to activate and on an underlying random variable  $\Theta$ called the state-of-the-world or the \textit{fundamental}. Herein, the structure of the $i$-th agent's payoff is the following function:
\begin{equation}\label{eq:payoff}
\mathcal{U}_i(a_i,a_{-i},\theta)\Equaldef a_i\Big(\sum_{j\neq i}a_j-\theta\Big), \ \ i\in[N].
\end{equation}
This payoff function is strictly increasing in the number of agents that activate for all $\theta$, and strictly decreasing in the fundamental for a fixed number of agents that activate. In other words, this game exhibits \textit{strategic complementarity}, i.e., when the decisions of the agents mutually reinforce one another, leading to the coordination of a particular behavior.  

The fundamental $\Theta$ is unknown and is randomly distributed on the positive real line $\mathbb{R}_+$. In the first part of the paper we assume that $\Theta$ is a binary random variable taking values on the set $\{\theta_0,\theta_1\}$, where $\theta_0<\theta_1$. In the context of QS, one possible interpretation for the fundamental is when $\theta_0$ denotes that a bacterial colony is in low concentration, and $\theta_1$ denotes that the colony is in high concentration; A second interpretation is when the fundamental represents the strength of the host organism's immune system: $\theta_0$ when it is weak and $\theta_1$ when it is strong. 
We assume that the prior probabilities are known:
\begin{equation}
 p_0 \Equaldef \mathbf{P}(\Theta= \theta_0) \ \ \text{and} \ \ p_1\Equaldef \mathbf{P}(\Theta= \theta_1).
\end{equation}
In the second part of the paper, we follow the literature on global games, by assuming that the fundamental is a continuous random variable distributed according to an (uninformative) uniform prior, i.e.,
\begin{equation}
f_{\Theta}(\theta) \Equaldef \begin{cases}
1, & \ \ \theta>0 \\
0, & \ \ \text{otherwise.}
\end{cases}
\end{equation}
This assumption is standard in global games and the fact that this is a degenerate distribution does not cause any technical difficulties. As we will see, all of our technical results rely on conditional distributions, all of which are well-defined.

%with $p<1/2$, without loss of generality.

%consisting a single species. Let $N$ be the number of cells in the colony. Each cell is an \textit{agent} in the game. 

%This number is assumed to be fixed and common knowledge among all agents. 

%Let $\Theta$ be the density of bacteria in the colony, defined as 
%\begin{equation}
%    \Theta = \frac{N}{V},
%\end{equation}
%where $V$ is the volume occupied the by $N$ agents. The quantity $\Theta$ is the  \textit{unknown} state-of-the-world. %In the parlance of the global game literature, $\Theta$ is also known as the \textit{fundamental}. 
%This quantity plays an important role in QS because, the decision of expressing a certain gene is only beneficial if $\Theta$ is high. %Let the volume be a random variable supported on the the positive real line $\mathbb{R_+}$ according to an arbitrary probability density function $f_V$.

As described in \cite{Vasconcelos2018BacterialSystem}, bacteria sense the environment via molecules, which are commonly modeled using Poisson processes. In our one-shot Bayesian game formulation, the observations are distributed according to Poisson a distribution with rate $\lambda$ modulated by the fundamental $\Theta$.  
Each agent receives a random private signal about the fundamental $\Theta$ distributed according to a Poisson distribution, i.e., the $i$-th agent observes
\begin{equation}
    Y_i  \sim \mathcal{P}(\lambda\Theta), \ \ i\in [N],
\end{equation}
where:
\begin{equation}
\mathbf{P}(Y_i=k\mid \Theta=\theta) = \frac{(\lambda\theta)^k}{k!}e^{-\lambda\theta}, \ \ k \in \mathbb{Z}_+.
\end{equation}
Finally, we assume that the collection of random variables $\{Y_i\}_{i=1}^N$ are conditionally independent given $\Theta$.  

The final ingredient of our game formulation are the policies. The $i$-th agent's action $a_i$ is computed according to a policy $\gamma_i:\mathbb{Z}_+ \rightarrow \{0,1\}$ such that
\begin{equation}
    a_i = \gamma_i(k) \ \ i \in[N].
\end{equation}
The goal of each agent is to maximize its expected pay-off function with respect to its policy $\gamma_i$, i.e.,
\begin{equation}
    \mathcal{J}_i(\gamma_i,\Gamma_{-i}) \Equaldef \mathbf{E}\Big[ \mathcal{U}_i\big(A_i,\sum_{j\neq i}A_j,\Theta\big) \Big].
\end{equation}

%\begin{equation}
%    \mathcal{U}_i(\mathbf{a}) = a_i\Bigg(\frac{b}{|\mathcal{N}_i|}\cdot \sum_{j\in \mathcal{N}_i}a_j \cdot \mathbf{1}(\Theta=\theta_1) - c \Bigg),
%\end{equation}
%where $b$ and $c$ are constants such that $b>c$, and which denote the \textit{benefit} and \textit{cost}, respectively,

%\begin{remark}
%The interpretation of the payoff function is directly inspired by the QS literature and is as follows: the expression of a gene is always costly, and the benefit only occurs if the colony is in the high concentration regime. Moreover, the benefit for activation is increasing in the number of neighboring bacteria that decide to activate. Notice that this payoff, is linearly increasing in the number of active neighbors. In practice, the benefit is increasing but bounded such that, the marginal benefit when the number of active neighbors is high has diminishing returns to scale.
%\end{remark}

\section{Best response to threshold strategies}

In the remainder of the paper, we will establish that for the global game with Poisson observations described in the previous section, the $i$-th agents best response to a strategy profile where all other agents use threshold policies, is itself a threshold policy. Then, we obtain sufficient conditions under which there exists a BNE in pure strategies.

We begin by defining the class of threshold policies, which will be the centerpiece of this paper.

\vspace{5pt}

\begin{definition}[Threshold policy]
A policy for the $i$-th agent is a threshold policy parameterized by a threshold $\tau_i \in \mathbb{Z}_+$ if it is of the following form:
\begin{equation}
    \gamma_i(k) = \begin{cases}
    1 & \text{if} \ \ k\leq \tau_i\\
    0 & \text{otherwise.}
    \end{cases}.
\end{equation}
We express the threshold policy in a more compact form by using an indicator function as $\gamma_i(k) = \mathbf{1}(k\leq \tau_i).$
\end{definition}

\vspace{5pt}

Let the collection of policies $\Gamma\Equaldef(\gamma_1,\cdots,\gamma_N)$ denote a  strategy profile for the agents in the game, and $\Gamma_{-i}\Equaldef(\gamma_1,\cdots,\gamma_{i-1},\gamma_{i+1},\cdots,\gamma_N)$ denote the strategies used by all the oponents of the $i$-th agent. 

\vspace{5pt}

\begin{definition}[Nash-Equilibrium]
A strategy profile $\Gamma^\star$ is a Nash-equilibrium if
\begin{equation}
 \mathcal{J}_i(\gamma^\star_i,\Gamma^\star_{-i}) \geq  \mathcal{J}_i(\gamma_i,\Gamma^\star_{-i}), \ \ i\in[N].
\end{equation}
\end{definition}

\vspace{5pt}

For an arbitrarily fixed strategy profile $\Gamma_{-i}$, we define the \textit{best-response} of the $i$-th agent to $\Gamma_{-i}$ as:
\begin{equation}
    \mathcal{R}^\star_{\Gamma_{-i}}(y) \Equaldef \arg \max_{\xi\in\{0,1\}} \mathbf{E}\Big[\mathcal{U}_i\big(\xi,\sum_{j\neq i}\gamma_j(Y_j),\Theta\big)\ \big| \ Y_i=y \Big].
\end{equation}
Using the structure of the payoff function in \cref{eq:payoff}, we obtain:
\begin{equation}\label{eq:best_response}
\mathcal{R}^\star_{\Gamma_{-i}}(k) = \begin{cases} 1, & \ \
\text{if} \ \ \pi_i(k) \geq \varphi(k) \\
 0, & \ \
\text{otherwise,}% \ \ \mathbf{E}\big[  \sum_{j\neq i}\gamma_j(Y_j) - \Theta  \mid Y_i=k \big] > 0.
\end{cases}
\end{equation}
where
\begin{equation}\label{eq:belief_threshold}
\pi_i(k) \Equaldef \mathbf{E} \Big[  \sum_{j\neq i} \gamma_j(Y_j) \mid Y_i=k \Big],
\end{equation}
and
\begin{equation}
\varphi(k) \Equaldef \mathbf{E}\big[\Theta \mid Y_i=k\big].
\end{equation}
Finally, when the strategy profile $\Gamma_{-i}$ consists only of threshold strategies, \cref{eq:belief_threshold} admits a convenient expression:
\begin{equation}\label{eq:belief}
\pi_i(k) = \sum_{i\neq j} \mathbf{P}(Y_j\leq \tau_j \mid Y_i=k).
\end{equation}

%Therefore, we have a decision rule of the following form:
%\begin{equation}
%\mathcal{R}^\star_{\Gamma_{-i}}(k) = \begin{cases} 0, & \ \
%\text{if} \ \ \mathbf{E}\big[  \sum_{j\neq i}\gamma_j(Y_j) - \Theta  \mid Y_i=k \big] \leq 0 \\
% 1, & \ \
%\text{if} \ \ \mathbf{E}\big[  \sum_{j\neq i}\gamma_j(Y_j) - \Theta  \mid Y_i=k \big] > 0.
%\end{cases}
%\end{equation}

\subsection{The binary state case}

\begin{proposition}\label{prop:fundamental_discrete}
Assume that $\Theta \in \{\theta_0,\theta_1\}$ such that $\theta_0<\theta_1$, and $p_0=\mathbf{P}(\Theta=\theta_0)$ and $p_1=\mathbf{P}(\Theta=\theta_1)$. Then
\begin{equation}\label{eq:conditional_mean}
\varphi(k) = \frac{\theta_0(\lambda\theta_0)^ke^{-\lambda\theta_0}p_0 + \theta_1(\lambda\theta_1)^ke^{-\lambda\theta_1}p_1 }{(\lambda\theta_0)^ke^{-\lambda\theta_0}p_0 + (\lambda\theta_1)^ke^{-\lambda\theta_1}p_1}
\end{equation}
and is strictly monotone increasing in $k$.
\end{proposition}

\vspace{5pt}

\begin{proof}
The function $\varphi(k)$ is computed using Bayes rule. To establish its strict monotonicity, we alternatively represent it as:
\begin{equation}
\varphi(k) = \frac{A + \theta_1p_1 \left(\frac{\theta_1}{\theta_0}\right)^k }{B + p_1 \left(\frac{\theta_1}{\theta_0}\right)^k}
\end{equation}
where
\begin{equation}
A \Equaldef e^{\lambda(\theta_1-\theta_0)}\theta_0p_0
\ \ 
\text{and}
\ \ 
B \Equaldef e^{\lambda(\theta_1-\theta_0)}p_0.
\end{equation}
We proceed by relaxing $k$ to be a real variable, and taking the partial derivative with respect to it. After some algebra, we show that the strict monotonicity condition $\varphi'(k)>0$ is equivalent to 
\begin{equation}
 \theta_1 > A/B = \theta_0.
\end{equation}
Since the $\varphi(k)$ is monotone increasing for $k\in\mathbb{R}_+$, it is also increasing for $k\in\mathbb{Z}_+$.
\end{proof}

\vspace{5pt}

The next step is to study the posterior belief of agent $i$ on the action  of agent $j$ conditioned on its observation $Y_i=k$, $j\neq i$. Consider the following conditional probability:
\begin{equation}
\pi_{ij}(k) \Equaldef \mathbb{P}(Y_j \leq \tau_j \mid Y_i = k).
\end{equation}
Computing this probability using Bayes' rule, we obtain:
\begin{equation}\label{eq:belief_discrete}
\pi_{ij}(k)=\sum_{m=0}^{\tau_j}\frac{(\lambda\theta_0)^{m+k}e^{-2\lambda\theta_0}p_0+(\lambda\theta_1)^{m+k}e^{-2\lambda\theta_1}p_1}{m!\Big[(\lambda\theta_0)^ke^{-\lambda\theta_0}p_0+(\lambda\theta_1)^ke^{-\lambda\theta_1}p_1\Big]}.
\end{equation}

\vspace{5pt}

We will show that the posterior belief in \cref{eq:belief_discrete} is strictly monotone decreasing in $k$. The proof of this result requires the following auxiliary lemma.

\vspace{5pt}

\begin{lemma}\label{lem:monotonicity_poisson}
Let
\begin{equation}
\mathcal{F}_{\tau} (\theta) \Equaldef \sum_{m=0}^\tau \frac{(\lambda\theta)^m}{m!} e^{-\lambda\theta}.
\end{equation}
The function $\mathcal{F}_\tau(\theta)$ is strictly monotone decreasing in $\theta$ for all $\tau \in \mathbb{Z}_+$.
\end{lemma}

\vspace{5pt}

\begin{proof}
The proof is in Appendix~\ref{ap:lemma}.
\end{proof}

\vspace{5pt}

\begin{proposition}\label{prop:belief_discrete}
Assume that $\Theta \in \{\theta_0,\theta_1\}$ such that $\theta_0<\theta_1$, and $p_0=\mathbf{P}(\Theta=\theta_0)$ and $p_1=\mathbf{P}(\Theta=\theta_1)$. Then $\pi_i(k) = \sum_{j\neq i} \pi_{ij}(k)$, where $\pi_{ij}(k)$ is given in \cref{eq:belief_discrete}, is strictly monotone decreasing in $k$.
\end{proposition}

\vspace{5pt}

\begin{proof}
Represent the conditional belief in \cref{eq:belief_discrete} as
\begin{equation}
\pi_{ij}(k) = \frac{h_j(k)}{g_j(k)},
\end{equation}
where 
\begin{equation}
%h_j(k) \Equaldef \sum_{m=0}^{\tau_j}\mathbf{E}\left[\frac{(\lambda\Theta)^{m+k}}{m!}e^{-2\lambda\Theta}\right].
h_j(k) \Equaldef \sum_{m=0}^{\tau_j}\frac{1}{m!}\Big(\lambda\theta_0)^{m+k}e^{-2\lambda\theta_0}p_0+(\lambda\theta_1)^{m+k}e^{-2\lambda\theta_1}p_1\Big)
\end{equation}
and 
\begin{equation}
g_j(k) \Equaldef (\lambda\theta_0)^ke^{-\lambda\theta_0}p_0+(\lambda\theta_1)^ke^{-\lambda\theta_1}p_1.
\end{equation}
Relaxing $k$ to be a real number, the strict monotone decreasing property is equivalent to:
\begin{equation}
\pi'_{ij}(k)<0 \Leftrightarrow h_j'(k)g_j(k) < h_j(k)g_j'(k),
\end{equation}
which, after some algebra, can be shown equivalent to:
\begin{equation}\label{eq:condition_discrete}
\sum_{m=0}^{\tau_j} \frac{(\lambda\theta_1)^m}{m!} e^{-\lambda\theta_1} <
\sum_{m=0}^{\tau_j} \frac{(\lambda\theta_0)^m}{m!} e^{-\lambda\theta_0}.
\end{equation}
Since $\theta_0<\theta_1$, \cref{lem:monotonicity_poisson} implies that the condition in \cref{eq:condition_discrete} is satisfied. Finally, since $\pi_{ij}(k)$
is monotone decreasing in $k$ for all $j$, $\pi_i(k)$ is also monotone decreasing.
\end{proof}

\vspace{5pt}

\begin{theorem}
Let $\Gamma_{-i}$ be a strategy profile consisting of threshold policies indexed by arbitrary thresholds. If $\Theta \in \{\theta_0,\theta_1\}$ is such that $\theta_0<\theta_1$, and $p_0=\mathbf{P}(\Theta=\theta_0)$ and $p_1=\mathbf{P}(\Theta=\theta_1)$, then $\mathcal{R}^\star_{\Gamma_{-i}}(k)$ is a threshold policy.
\end{theorem}

\vspace{5pt}

\begin{proof}
From \cref{prop:fundamental_discrete,prop:belief_discrete}, the conditional expectation of the fundamental $\Theta$ given the $i$-th agent's observation $Y_i=k$, is monotone increasing in $k$ and that the $i$-th agent's belief on the $j$-th agent's action is monotone decreasing in $k$ when it uses a threshold policy. Therefore, the aggregate belief on the actions of the remaining agents in the system is also monotone decreasing in $k$. The implication of these two facts is that there is at most a single crossing point $k^\star$ between the discrete functions $\pi_i(k)$ and $\varphi(k)$. Therefore, the best response policy $\mathcal{R}_i^\star(k)$ in \cref{eq:best_response} to any given threshold strategy profile $\Gamma_{-i}$ is a threshold strategy.
\end{proof}

%Therefore, the aggregate of the actions of all the $N-1$ agents is also monotone decreasing in $k$.
%\begin{equation}
%\sum_{j\neq i} \mathbf{E} \left[ \gamma_j(Y_j) \mid Y_i=k \right] = 
%\sum_{j\neq i} \pi_{ij}(k;\tau_j):
%\end{equation}

%This implies that the best response to a collection $\Gamma_{-i}$ of threshold policies characterized by $\tau_j \in \mathbb{N}$, such that $j\neq i$, is also a threshold policy characterized by $\tau_i^\star$, where:

\vspace{5pt}

\begin{remark}
Depending on the parameters of the problem, it is possible that the threshold that characterizes the best-response to a particular collection of threshold policies is infinite, leading to a \textit{degenerate} threshold policy. 
\end{remark}

%There may be a case where $\tau_i^{\star}$ is infinite. That case corresponds to the strict inequality below: 
%\begin{equation}
%\mathbf{E}[\Theta \mid Y_i=0] < \sum_{j\neq i}\pi_{ij}(0;\tau_j).
%\end{equation}

%However, in any other case, i.e., when 
%\begin{equation}
%\mathbf{E}[\Theta \mid Y_i=0] \geq \sum_{j\neq i}\pi_{ij}(0;\tau_j).
%\end{equation}

%There is exactly 1 crossing point between the two curves.
%which can be established by the assimptotic behavior of the two functions: \textcolor{magenta}{we may need to assume for completeness that $\theta_1>p_1$.}

%\begin{equation}
%\lim_{k\rightarrow \infty} \mathbf{E}[\Theta \mid Y_i=k] = \theta_1. 
%\end{equation}
%whereas,
%\begin{equation}
%\lim_{k\rightarrow \infty} \sum_{j\neq i} \pi_{ij}(k;\tau_j) = \sum_{j\neq i} \sum_{m=0}^{\tau_j} \frac{(\lambda\theta_1)^m}{m!}. 
%\end{equation}

%Showing the existence of a Nash Equilibrium in deterministic threshold strategies in this case may be impossible due to the non-continuity of the best response map and the fact that there policy space has infinite cardinality.

\vspace{5pt}

\begin{theorem}
For fixed $N$, $\lambda$, $\theta_0$, $\theta_1$, $p_0$ and $p_1$, such that 
\begin{equation}
\theta_1 > N-1,
\end{equation}
and
\begin{equation}\label{eq:lambda}
\mathbf{E}\left[\big((N-1)e^{-2\lambda \Theta}- \Theta e^{-\lambda \Theta}\big) \right] > 0, 
\end{equation}
there exists a BayesianNash-Equilibrium in the class of pure threshold strategies.
\end{theorem}

\vspace{5pt}

\begin{proof}To obtain a finite single crossing property, we start with the following condition:
\begin{equation}\label{eq:condition1}
\pi_i(0) < \varphi(0), \ \ i \in [N].
\end{equation}

Notice that
\begin{equation}
\pi_i(0) \geq (N-1)\left(\frac{e^{-2\lambda \theta_0}p_0 +e^{-2\lambda \theta_1}p_1 }{e^{-\lambda \theta_0}p_0 +e^{-\lambda \theta_1}p_1 }\right).
\end{equation}

and 
\begin{equation}
\varphi(0) = \left(\frac{\theta_0e^{-\lambda \theta_0}p_0 +\theta_1e^{-\lambda \theta_1}p_1 }{e^{-\lambda \theta_0}p_0 +e^{-\lambda \theta_1}p_1 }\right).
\end{equation}
Therefore, if the following inequality holds,
\begin{multline}\label{eq:sufficient_condition1}
(N-1)\big(e^{-2\lambda \theta_0}p_0 +e^{-2\lambda \theta_1}p_1 \big) > \theta_0e^{-\lambda \theta_0}p_0 +\theta_1e^{-\lambda \theta_1}p_1,
\end{multline}
the condition is \cref{eq:condition1} is met. Rearranging the terms in \cref{eq:sufficient_condition1}, we obtain the expression in the \cref{eq:lambda}. The second condition to obtain the finite single crossing property is:
\begin{equation}
\lim_{k\rightarrow \infty} \pi_i(k) < \lim_{k\rightarrow \infty} \varphi(k).
\end{equation}

Since,
\begin{equation}
\lim_{k\rightarrow \infty} \varphi(k) = \theta_1
\end{equation}
and
\begin{equation}
\lim_{k\rightarrow \infty} \pi_i(k) = \sum_{j\neq i} \sum_{m=0}^{\tau_j} \frac{(\lambda\theta_1)^m}{m!}e^{-\lambda \theta_1} \leq N-1.
\end{equation}
Thus, we obtain $\theta_1 > N-1$.

Then, if the two conditions are satisfied, the best-response to any threshold strategy profile is a threshold strategy with a finite threshold. Therefore, the agents in the system can choose threshold strategies from a finite set, turning this into a finite game. Finally, notice that the payoffs constitute a potential game, and every finite potential game has a Nash-equilibrium in pure strategies \cite{hespanha2017noncooperative}. 
\end{proof}

\vspace{5pt}

\begin{corollary}[The role of $\lambda$]\label{cor:role}
If 
\begin{equation}
\mathbf{E}[\Theta] < N-1,
\end{equation} there exists a small enough $\bar{\lambda}$ such that 
\begin{equation}
\mathbf{E}\Big[\big((N-1)e^{-2\bar{\lambda} \Theta}- \Theta e^{-\bar{\lambda} \Theta}\big) \Big] > 0
\end{equation}
holds.
\end{corollary}

\vspace{5pt}

\begin{proof}
The LHS of the inequality in \cref{eq:lambda} evaluated at $\lambda=0$ is equal to
\begin{equation}
N-1-\mathbf{E}[\Theta] > 0.
\end{equation}
The results follows from the continuity of the LHS of \cref{eq:lambda} in $\lambda$.
\end{proof}

\begin{remark}
\cref{cor:role} establishes a condition on the average fundamental and the number of agents in the system, such that there exists strictly positive value of $\lambda$ for the existence of BNE. One way to interpret this result is by imagining that the mechanism that generates the private signals $\{Y_i\}_{i=1}^\infty$ can be adjusted to guarantee that the a finite game is played by the agents. This may have implications in engineered distributed systems, where it is important that the agents optimize parameters within a finite set. If, for example, the agents are learning the optimal policy from repeated games such as in \textit{fictitious play} \cite{hespanha2017noncooperative}.
\end{remark}

\subsection{Numerical Example}\label{sec:example_discrete}
\begin{figure}[b!]
    \centering
    \includegraphics[width=\columnwidth]{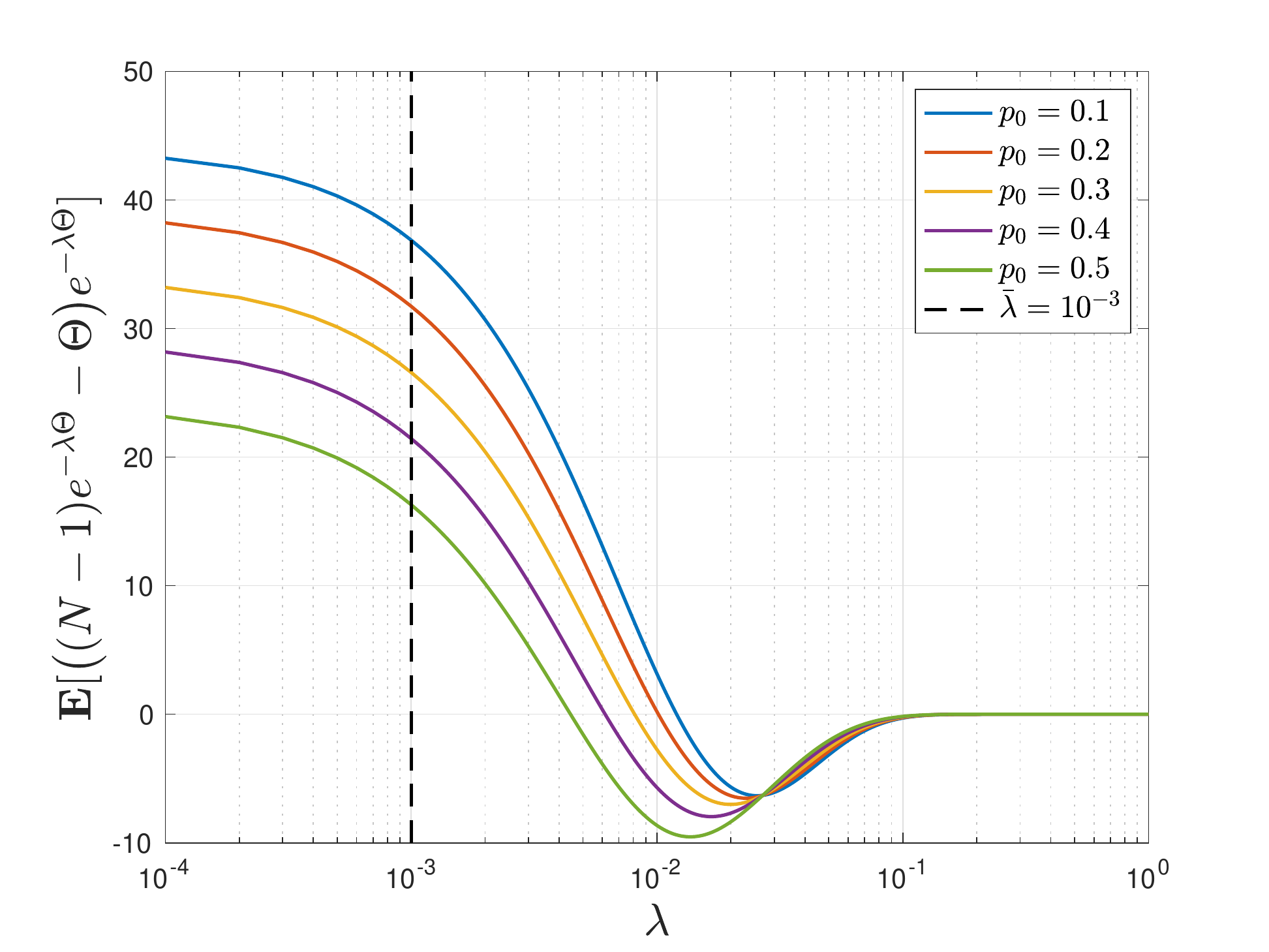}
    \caption{Condition in \cref{eq:condition_discrete} for the existence a single threshold.}
    \label{fig:existence}
\end{figure}

We consider a game with binary state and the following parameters: $N=100$ agents, $\theta_0=50$, $\theta_1=100$. From \cref{fig:existence}, we observe that for $\bar{\lambda}=10^{-3}$, the finite crossing property is satisfied for $p_0 \in \{0.1, 0.2, 0.3, 0.4, 0.5\}$. Running the best-response dynamics for any initial condition is guaranteed to converge to a Nash-equilibrium. In the case of this example, the following strategy profiles constitute Nash-equilibria:
\begin{equation}
\Gamma^\star = (\gamma^\star, \gamma^\star,\cdots, \gamma^\star),
\end{equation}
where
\begin{equation}
\gamma^\star(k) = \mathbf{1}(k\leq \tau^\star). 
\end{equation}
The value of $\tau^\star$ depends on $p_0$ and can be found on \cref{tab:my_label}.

\begin{table}
    \centering
    \caption{Threshold for the Nash-equilibrium strategies in \cref{sec:example_discrete}}
    \begin{tabular}{cc}
        $p_0$ & $\tau^\star$  \\
        \hline \hline
        $0.1$ & $5$ \\
        $0.2$ & $6$ \\
        $0.3$ & $6$ \\
        $0.4$ & $7$ \\
        $0.5$ & $8$ \\
        \hline \hline
    \end{tabular}
    \label{tab:my_label}
\end{table}

\section{Continuous fundamental with uniform distribution}

In the global game literature, the prior distribution on the fundamental $\Theta$ is commonly assumed to be the so-called \textit{uninformative uniform distribution} \cite{Touri2014GlobalInformation,mahdavifar2017global,Dahleh:2016}. Since the fundamental in our game formulation controls the rate of a Poisson distribution, it only makes sense to consider the following (degenerate) distribution on the positive real numbers, i.e.,
\begin{equation}
f_{\Theta}(\theta) = 1, \ \ \theta>0.
\end{equation}

We begin our analysis by computing the posterior distribution on the fundamental given a Poisson observation. Unlike the canonical global game with Gaussian observations and (degenerate) uniformly distributed fundamental, in which the posterior belief is Gaussian, the Poisson observation model results in a Gamma posterior distribution. To see this, let $Y_i\sim \mathcal{P}(\lambda\Theta)$, then\footnote{Notice this is a degenerate distribution as well. However, the conditional density (belief) is well-defined.}
\begin{equation}
\mathbf{P}(Y_i=k) = \int_{0}^{\infty} \frac{(\lambda\theta)^k}{k!}e^{-\lambda\theta}d\theta = \frac{1}{\lambda}, \ \ k\in \mathbb{Z}_+.
\end{equation}
Therefore, the conditional belief on the fundamental is given by:
\begin{equation}
f_{\Theta \mid Y_i = k}(\theta) = \frac{\mathbf{P}(Y_i=k \mid \Theta = \theta)f_{\Theta}(\theta)}{\mathbf{P}(Y_i=k)}, 
\end{equation}
which is equal to:
\begin{equation}
f_{\Theta \mid Y_i = k}(\theta) = \begin{cases}
\lambda \frac{(\lambda\theta)^k}{k!}e^{-\lambda\theta}, & \ \ \textrm{if} \ \  \theta>0 \\
0, & \ \ \textrm{otherwise.}
\end{cases} 
\end{equation}
This is a well-defined conditional probability distribution on $\Theta$. In fact, $f_{\Theta \mid Y_i = k}$ is a particular instance of a Gamma distribution with shape $k+1$ and scale $1/\lambda$. Of paramount importance to our problem is the mean of this distribution, i.e.,
\begin{equation}\label{eq:mean_fundamental}
\mathbf{E} \big[\Theta \mid Y_i=k\big] = \frac{k+1}{\lambda}\Equaldef  \varphi(k) ,
\end{equation}
which is clearly a monotone increasing function of $k$. This property bears a lot of similarity with the global game literature with Gaussian observations. 

We proceed with the analysis of the $i$-th agent's belief on the action of the other agent $j$, $i\neq j$. Define the belief of agent $i$ on the action of agent $j$ as:
\begin{equation}
\pi_{ij}(k) \Equaldef \mathbf{P} (A_j=1 \mid Y_i =k), \ \ k=0,1,\cdots
\end{equation}

\vspace{5pt}

\begin{proposition}\label{prop:belief}
Assume that the $j$-th agent uses a threshold policy $\gamma_j(m) = \mathbf{1}(m\leq \tau_j)$. Then,
\begin{equation}\label{eq:belief_continuous}
    \pi_{ij}(k)=\sum_{m=0}^{\tau_j} 2^{-(m+k+1)}\binom{k+m}{k}.
\end{equation}
\end{proposition}
\vspace{5pt}
\begin{proof}
From the definition of $\pi_{ij}$ and the assumption that agent $j$ uses a threshold policy, we have:
\begin{equation}
\pi_{ij} (k) = \sum_{m=0}^{\tau_j} \mathbf{P}(Y_j=m \mid Y_i=k).
\end{equation}
The following equalities hold:
\begin{IEEEeqnarray}{rCl}
\mathbf{P}(Y_j&=&m \mid Y_i=k)  \\
& = & \int_{0}^{\infty} \mathbf{P}(Y_j=m,\Theta=\theta \mid Y_i = k)d\theta \\
& = & \int_{0}^{\infty} \mathbf{P}(Y_j=m\mid \Theta=\theta) f_{\Theta \mid Y_i = k}(\theta)d\theta \\
%& = & \int_{0}^{\infty} \frac{(\lambda\theta)^m}{m!}e^{-\lambda\theta} \lambda \frac{(\lambda\theta)^k}{k!}e^{-\lambda\theta}d\theta \\
& = & \frac{\lambda}{m!\cdot k!}\int_{0}^{\infty} (\lambda\theta)^{(k+m)}e^{-2\lambda\theta} d\theta \\
%& = & \frac{\lambda}{m! \cdot k!}\cdot\frac{(m+k)!}{\lambda\cdot 2^{m+k+1}} \\
& = &  2^{-(m+k+1)}\binom{k+m}{k}.
%\int_{0}^{\infty} (\lambda\theta)^{(m+k)}e^{-2\lambda\theta} d\theta \\
\end{IEEEeqnarray}
\end{proof}

\vspace{5pt}

\begin{remark}
Notice that, unlike the binary fundamental case, this belief $\pi_{ij}$ is independent of the rate parameter $\lambda$. 
\end{remark}

\vspace{5pt}

\begin{remark}
A consequence of \cref{prop:belief} is the following combinatorial identity:
\begin{equation}\label{eq:identity}
\sum_{m=0}^{\infty} 2^{-(m+1)}\binom{k+m}{k} = 2^k.
\end{equation}
This identity is not trivial and will be used in the proof of \cref{lem:monotone_belief}. 
\end{remark}

\vspace{5pt}

The next step in our argument is to show that the belief $\pi_{ij}(k)$ is a monotone decreasing function of $k$ for all $\tau_j$. %The figure below shows a numerical evidence that such claim is true. However, the complete proof of this claim is a difficult problem in discrete mathematics and possibly combinatorics, to which we present a partial result.

%\begin{figure}[h!]
%    \centering
%    \includegraphics[width=0.8\columnwidth]{belief.pdf}
%    \caption{Belief of agent $i$ on the activation of agent $j$ assuming a threshold policy with $\tau_j$.}
%    \label{fig:belief}
%\end{figure}

\vspace{5pt}

\begin{lemma}\label{lem:monotone_belief}
The belief $\pi_{ij}(k)$ in \cref{eq:belief_continuous} is monotone decreasing in $k$ for all $\tau_j \in \mathbb{Z}_+$.
\end{lemma}

\vspace{5pt}

\begin{proof}
To prove the monotonicity of $\pi_{ij}$, we define the first order difference $\Delta_{\tau}(k)$ and show that it is strictly positive for all $k$. Let
\begin{equation}\label{eq
:first_order}
\Delta_{\tau}(k)\Equaldef \pi_{ij}(k) - \pi_{ij}(k+1). 
\end{equation}
The following equalities hold:
\begin{align}
 &\Delta_{\tau}(k)\nonumber \\
 &=\sum_{m=0}^{\tau} \left[ \binom{k+m}{k} - \frac{1}{2}\binom{k+1+m}{k+1}\right]2^{-(k+m+1)} \\
 & \stackrel{(a)}{=}  \sum_{m=0}^{\tau} \left[ \binom{k+m}{k} - \binom{k+m}{k+1}\right]2^{-(k+m+2)} \\
 & =  \sum_{m=0}^{\tau} \left[ \frac{k-m+1}{m+k+1}\right]\binom{m+k+1}{k+1}2^{-(k+m+2)} \\
 & = \sum_{m=k+1}^{\tau} \left[ \frac{k-m+1}{m+k+1}\right]\binom{m+k+1}{k+1}2^{-(k+m+2)} \nonumber \\
 &  + \sum_{m=0}^{\min\{k,\tau\}} \left[ \frac{k-m+1}{m+k+1}\right]\binom{m+k+1}{k+1}2^{-(k+m+2)}, \label{eq:two_cases}
\end{align}
where $(a)$ follows from Pascal's triangle\footnote{The indentity known as Pascal's triangle (or formula) is:$$\binom{n}{k}=\binom{n-1}{k-1}+\binom{n-1}{k}, \ \ \ 1\leq k\leq n.$$}. Based on Eq. \eqref{eq:two_cases}, we need to consider two cases: %$1)$ $k\geq\tau$ and $2)$ $k<\tau$.

\vspace{5pt}

\noindent \textbf{Case 1: $(k\geq \tau)$}
 Notice that in this regime, we have:
\begin{equation}
\Delta_{\tau}(k)=\sum_{m=0}^{k} \left[ \frac{k-m+1}{m+k+1}\right]\binom{m+k+1}{k+1}2^{-(k+m+2)}
\end{equation}
and all of the terms in the summation are positive, which immediately implies that:
\begin{equation}
\Delta_{\tau}(k) > 0.
\end{equation}

\noindent \textbf{Case 2:} $(k < \tau)$
In this regime, the first order difference in \cref{eq
:first_order} can be written as:
\begin{equation}
\Delta_{\tau}(k) = A_{\tau}(k) + B(k),
\end{equation}
where
\begin{equation}
A_{\tau}(k) \Equaldef \sum_{m=k+1}^{\tau} \left[ \frac{k-m+1}{m+k+1}\right]\binom{m+k+1}{k+1}2^{-(k+m+2)},
\end{equation}
and
\begin{equation}
B(k) \Equaldef \sum_{m=0}^{k} \left[ \frac{k-m+1}{m+k+1}\right]\binom{m+k+1}{k+1}2^{-(k+m+2)}.
\end{equation}
Then, notice that for a fixed $k$, the function $A_{\tau}(k)$ is monotone decreasing in $\tau$ because the arguments in the summation are negative. Thus, the follwoing strict inequality holds:
\begin{equation}\label{eq:bound}
\Delta_\tau(k) > A_{\infty}(k) + B(k) , \ \ k\in\mathbb{Z}_+.
\end{equation}
We will show that the RHS of \cref{eq:bound} is identically equal to zero.
\begin{align}
&A_\infty(k) + B(k) \nonumber \\ & =  \sum_{m=0}^{\infty} \left[ \frac{k-m+1}{m+k+1}\right]\binom{m+k+1}{k+1}2^{-(k+m+2)} \\
& =   \sum_{m=0}^{\infty} \left[ 1 - \frac{m}{k+1}\right]\binom{m+k}{m}2^{-(k+m+2)} \\
& =  \frac{1}{2}\sum_{m=0}^{\infty} \binom{m+k}{m}2^{-(k+m+1)} \nonumber \\ & \qquad \qquad \qquad - \frac{1}{2}\sum_{m=0}^{\infty} \binom{m+k+1}{m}2^{-(k+m+2)} \\
& \stackrel{(b)}{=} \frac{1}{2} - \frac{1}{2} =0,
\end{align}
where $(b)$ follows from \cref{eq:identity}.
\end{proof}

\vspace{5pt}

\begin{theorem}
Let $\Gamma_{-i}$ be a strategy profile consisting of threshold policies indexed by arbitrary thresholds. If $\Theta$ is uniformly distributed on $\mathbb{R}_+$, then $\mathcal{R}^\star_{\Gamma_{-i}}(k)$ is a threshold policy.
\end{theorem}

\vspace{5pt}

\begin{proof}
From \cref{lem:monotone_belief}, the functions $\pi_{ij}(k)$ are monotone decreasing for all $i,j \in [N]$, and $i\neq j$. Therefore, $\pi_i(k)$ in \cref{eq:belief} is monotone decreasing. Since $\varphi(k)$ in \cref{eq:mean_fundamental} is strictly increasing, there is at most a single point at which the two sequences cross. Hence, the best response policy $\mathcal{R}_i^\star(k)$ to any given threshold strategy profile $\Gamma_{-i}$ is a threshold strategy.
\end{proof}

\subsection{Existence of Nash-equilibria}

\begin{theorem}
If the rate $\lambda > 2/(N-1)$, then there exists a Nash-Equilibrium in the class of pure threshold strategies.
\end{theorem}

\vspace{5pt}

\begin{proof}
Recall that the best-response to a threshold strategy profile $\Gamma_{-i}$ is given by:
\begin{equation}
\sum_{j\neq i} \pi_{ij}(k) \underset{\mathcal{R}^\star(\Gamma_{-i})(k)=1 }{\overset{\mathcal{R}^\star(\Gamma_{-i})(k)=0 }{\lesseqgtr}} \frac{k+1}{\lambda}.
\end{equation}
\Cref{lem:monotone_belief} implies that the LHS of the inequality above is monotone decreasing in $k$, whereas the RHS is clearly monotone increasing. The function on the RHS is equal to $1/\lambda$ at $k=0$. On the otherhand, at $k=0$, the function on the LHS can be lower bounded by 
\begin{equation}
\sum_{j\neq i} \big(1-2^{-(\tau_j+1)}\big) \in \left[\frac{N-1}{2},N-1 \right).
\end{equation}
Therefore, if 
\begin{equation}
\frac{1}{\lambda} < \frac{N-1}{2},
\end{equation}
there is a single solution to:
\begin{equation}
\tau_i^\star = \max \Big\{ k \in \mathbb{Z}_+ \mid \sum_{j\neq i}\pi_{ij}(k) > \frac{k+1}{\lambda} \Big\}.
\end{equation}

Therefore, the agents in the system can choose threshold strategies from a finite set without loss of optimality, turning this into a finite game. Finally, notice that the payoffs constitute a potential game, and every finite potential game has a Nash-equilibrium in pure strategies \cite{hespanha2017noncooperative}. 

\end{proof}

%\begin{remark}
%This result is important because for the problem with a binary fundamental, the condition for such existence is either unavailable or does not exist in closed form.
%\end{remark}

%\begin{equation}
%    \pi_{ij}(k)=\frac{1}{2}\sum_{\ell=\tau_j}^\infty
%    \frac{\binom{k+\ell}{k}}{2^{\ell+k}}
%\end{equation}

%
%\begin{equation}
%    \Pr(|\mathcal{S}|\leq k)
%\end{equation}

\subsection{Numerical example: multiplicity of equilibria}

\begin{figure}[b!]
    \centering
    \includegraphics[width=\columnwidth]{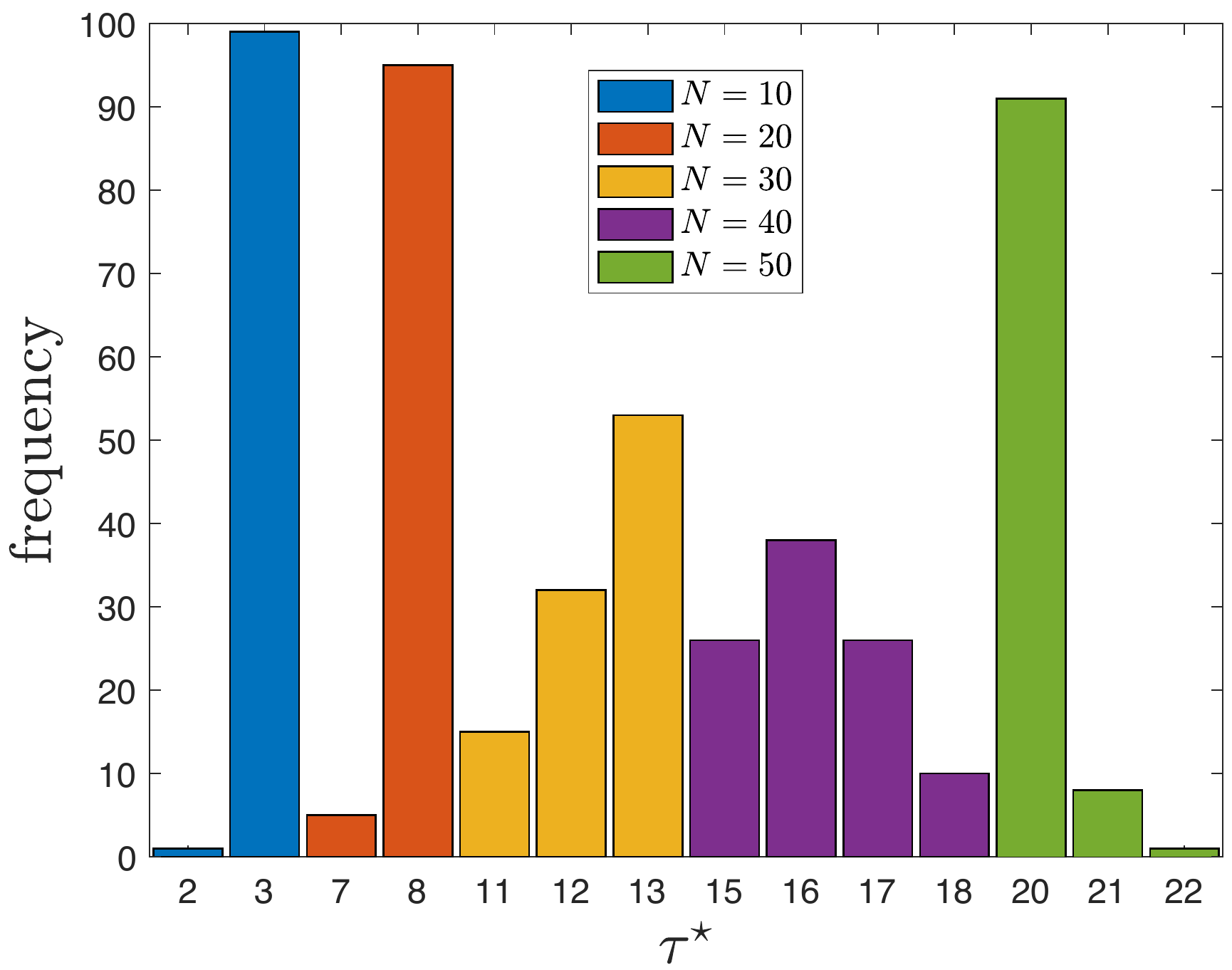}
    \caption{Thresholds for the BNE strategy profile $\Gamma^\star$ obtained using best-response dynamics and their frequency within $M=100$ experiments.}
    \label{fig:multiplicity}
\end{figure}

Consider the global game with uniform prior with $\lambda = 1$, and $N=10, 20, 30, 40$, and $50$ agents. Randomly initializing the threshold at every agent according to a Geometric random variable with parameter $p=0.05$, we repeatedly compute the best-response strategies until convergence. This process was repeated $M=100$ times and the threshold $\tau^\star$ corresponding to the Nash equilibrium was recorded. This numerical experiment has revealed the existence of multiple equilibria, which can be visualized in \cref{fig:multiplicity}. For the Gaussian counterpart of this game, the equilibrium is unique, unveiling a potential fragility of Poisson signaling for coordination of bacterial populations via QS.

Finally, an equilibrium for a game with $N=100$ agents was computed using best-response dynamics, and the threshold obtained was $\tau^\star =44$ for all policies in $\Gamma^\star.$ Since the convergence of the best-response dynamics for a system with $N=100$ agents is very slow, we were not able to repeat it a significant number of times. We conjecture that as the number of agents increases, the equilibrium in pure threshold strategies becomes unique. Such an issue is an open research problem for future work.

%One implication of the Theorems in the previous sections is that, the global game over the class of threshold strategies is well defined, and the structural results herein allow us to view them as matrix games, where the actions are the thresholds. In the first case, we do not presently have a sufficient condition that guarantees the existence of a threshold for any strategy profile. Therefore, the game has an infinite action space, which violates one of the cannonical conditions for the existence of Nash-Equilibria. On the other hand, in the continous fundamental with uniformly distributed prior, we do have a sufficient condition for existence of a best-response threshold, but since the game takes place in a discrete space, to which we do not know the cardinality a priori, we cannot guarantee the existence of a Nash-Equilibirum in mixed strategies either.

%However, we follow the literature of Learning in games to show that... 

%We will not address the issue of learning the equilibria here.

%Instead, we will establish the existence of a Nash-equilibria in threshold policies by showing that this game is a finite Potential game. Therefore, it admits an equilibrium in pure strategies.

%Then, by means of an example, we will discuss the existence of multiple equilibria.

\section{Conclusions and Future Work}

Global games have been widely used to model economic and social mechanisms where the agents have the incentive to coordinate their actions. We argue that such element of coordination is also present in many biological phenomena, of which bacterial quorum-sensing is an important example. Therefore, global games may be an appropriate framework to study the coordination of bacterial behavior. However, unlike the Gaussian signals considered in the existing literature, bacteria sense their environment using molecular signals, which are commonly modeled using a Poisson distribution. 

Here, we studied the first examples of global games with Poisson observations. When the fundamental is binary or uniform continuous, we have proved that the best-response policy to a strategy profile consisting of threshold policies is itself a threshold policy. Moreover, we have also provided sufficient conditions for the existence of equilibria in pure threshold strategies for such games.

Future work will focus on learning the optimal threshold via fictitious play, and its rate of convergence; the existence of equilibria for more general payoff functions; and the introduction of agents of a second type whose role is to inhibit the coordination of the agents of first type.

\appendices

\section{Proof of \cref{lem:monotonicity_poisson}}\label{ap:lemma}
\begin{proof}
We prove the result in three steps.
Let $\tau=0$, then:
\begin{equation}
\mathcal{F}_0'(\theta)=(e^{-\lambda\theta})' = -\lambda e^{-\lambda\theta}<0. 
\end{equation}
Let $\tau=1$, then:
\begin{equation}
\mathcal{F}_1'(\theta) = (e^{-\lambda\theta} + \lambda\theta e^{-\lambda\theta})'= -\lambda^2\theta e^{-\lambda\theta} <0.
\end{equation}

We can now prove the general case when $\tau\geq 2$:
\begin{IEEEeqnarray}{rCl}
\mathcal{F}'_\tau(\theta) & = &  \Big(\sum_{k=0}^{\tau} \frac{(\lambda\theta)^k}{k!}e^{-\lambda\theta}\Big)' \\
& = & \Big(e^{-\lambda\theta} + \lambda\theta e^{-\lambda\theta}+\sum_{k=2}^{\tau-1} \frac{(\lambda\theta)^k}{k!}e^{-\lambda\theta}\Big)' \\
& = &  \lambda \sum_{k=2}^{\tau-1} \frac{(\lambda\theta)^{k}}{k!} e^{-\lambda\theta} - \lambda \sum_{k=2}^{\tau} \frac{(\lambda\theta)^{k}}{k!} e^{-\lambda\theta}\\
& = & -\lambda \frac{(\lambda\theta)^{\tau}}{\tau!}e^{-\lambda\theta} <0.
\end{IEEEeqnarray}
Therefore,
\begin{equation}
\mathcal{F}'_\tau(\theta) < 0, \ \ \tau\in\mathbb{Z}_+.
\end{equation}
\end{proof}

\bibliographystyle{IEEEtran}
    \bibliography{IEEEabrv,references}

\begin{thebibliography}{10}
\providecommand{\url}[1]{#1}
\csname url@samestyle\endcsname
\providecommand{\newblock}{\relax}
\providecommand{\bibinfo}[2]{#2}
\providecommand{\BIBentrySTDinterwordspacing}{\spaceskip=0pt\relax}
\providecommand{\BIBentryALTinterwordstretchfactor}{4}
\providecommand{\BIBentryALTinterwordspacing}{\spaceskip=\fontdimen2\font plus
\BIBentryALTinterwordstretchfactor\fontdimen3\font minus
  \fontdimen4\font\relax}
\providecommand{\BIBforeignlanguage}[2]{{%
\expandafter\ifx\csname l@#1\endcsname\relax
\typeout{** WARNING: IEEEtran.bst: No hyphenation pattern has been}%
\typeout{** loaded for the language `#1'. Using the pattern for}%
\typeout{** the default language instead.}%
\else
\language=\csname l@#1\endcsname
\fi
#2}}
\providecommand{\BIBdecl}{\relax}
\BIBdecl

\bibitem{Carlsson1993GlobalSelection}
H.~Carlsson and E.~van Damme, ``Global games and equilibrium selection,''
  \emph{Econometrica}, vol.~61, no.~5, 1993.

\bibitem{Morris2003GlobalApplications}
S.~Morris and H.~S. Shin, ``{Global games: Theory and applications},'' in
  \emph{Advances in Economics and Econometrics: Theory and Applications, Eighth
  World Congress, Volume 1}, 2003.

\bibitem{Krishnamurthy2008DecentralizedGames}
V.~Krishnamurthy, ``{Decentralized activation in dense sensor networks via
  global games},'' \emph{IEEE Transactions on Signal Processing}, vol.~56,
  no.~10, 2008.

\bibitem{krishnamurthy2011networks}
------, ``Networks of biosensors: decentralized activation and social
  learning,'' \emph{European journal of control}, vol.~17, no. 5-6, pp.
  526--546, 2011.

\bibitem{Kanakia2016ModelingGame}
A.~Kanakia, B.~Touri, and N.~Correll, ``{Modeling multi-robot task allocation
  with limited information as global game},'' \emph{Swarm Intelligence},
  vol.~10, no.~2, 2016.

\bibitem{Miller:2001}
M.~B. Miller and B.~L. Bassler, ``Quorum sensing in bacteria,'' \emph{Annual
  Review of Microbiology}, vol.~55, no.~1, pp. 165--199, 2001.

\bibitem{Darch2012Density-dependentPopulations}
S.~E. Darch, S.~A. West, K.~Winzer, and S.~P. Diggle, ``{Density-dependent
  fitness benefits in quorum-sensing bacterial populations},''
  \emph{Proceedings of the National Academy of Sciences}, vol. 109, no.~21,
  2012.

\bibitem{Boedicker2015MicrobialSensing}
J.~Boedicker and K.~Nealson, ``{Microbial communication via quorum sensing},''
  \emph{IEEE Transactions on Molecular, Biological, and Multi-Scale
  Communications}, vol.~1, no.~4, 2015.

\bibitem{Michelusi2016QueuingCommunities}
N.~Michelusi, J.~Boedicker, M.~Y. El-Naggar, and U.~Mitra, ``Queuing models for
  abstracting interactions in bacterial communities,'' \emph{IEEE Journal on
  Selected Areas in Communications}, vol.~34, no.~3, 2016.

\bibitem{Vasconcelos2018BacterialSystem}
M.~M. Vasconcelos, U.~Mitra, O.~Camara, K.~P. Silva, and J.~Boedicker,
  ``Bacterial quorum sensing as a networked decision system,'' in \emph{IEEE
  International Conference on Communications}, 2018.

\bibitem{Fudenberg1991Game1991}
D.~Fudenberg and J.~Tirole, \emph{Game theory}.\hskip 1em plus 0.5em minus
  0.4em\relax The MIT press, 1991.

\bibitem{chamley2004rational}
C.~P. Chamley, \emph{Rational herds: Economic models of social learning}.\hskip
  1em plus 0.5em minus 0.4em\relax Cambridge University Press, 2004.

\bibitem{Dahleh:2016}
M.~A. Dahleh, A.~{Tahbaz-Salehi}, J.~N. Tsitsiklis, and S.~I. Zampoulis,
  ``{Coordination with local information},'' \emph{Operations Research},
  vol.~64, no.~3, 2016.

\bibitem{Touri2014GlobalInformation}
B.~Touri and J.~Shamma, ``Global games with noisy sharing of information,'' in
  \emph{Proceedings of the IEEE Conference on Decision and Control}, 2014.

\bibitem{mahdavifar2017global}
H.~Mahdavifar, A.~Beirami, B.~Touri, and J.~S. Shamma, ``Global games with
  noisy information sharing,'' \emph{IEEE Transactions on Signal and
  Information Processing over Networks}, vol.~4, no.~3, pp. 497--509, 2017.

\bibitem{hespanha2017noncooperative}
J.~P. Hespanha, \emph{Noncooperative Game Theory}.\hskip 1em plus 0.5em minus
  0.4em\relax Princeton University Press, 2017.

\end{thebibliography}

\end{document}